# Predicting and increasing subjective well-being: response function model and rhythmic movement therapy

**Irina G. Malkina-Pykh**
Research Center for Interdisciplinary Environmental Cooperation
Russian Academy of Sciences (INENCO RAS)
Saint-Petersburg, Russia
Nab. Kutuzova, 14
+7(812)2721601
+7(961)8099247
e-mail: inenco@mail.neva.ru
malkina@mail.admiral.ru
www.inenco.org

**Abstract**

**Background.** The objective of the present study was to apply the nonlinear response function model of subjective well-being (RFSWB model) to evaluate the outcome of rhythmic movement therapy (RMT) for increasing subjective well-being and to analyze whether intervention-related changes in several psychological variables were mechanisms underlying SWB increase in subjects participating in RMT group. **Methods.** A total of 273 subjects (54 males and 219 females, mean age was 37.3±10.5 years) were selected at random in nonclinical population and assessed with the appropriate surveys and questionnaires. The RMT program was proposed to the 105 subjects (24 males, 81 females, and mean age 37.6±11.7 years) with very low, low and medium SWB level. Control group was included. **Findings.** Results revealed that: a) substantial changes in SWB and underlying psychological state were observed among the participants as a result of RMT intervention; b) RFSWB model predicts the changes in SWB after RMT intervention satisfactorily and can help to identify the reliable predictors of success.

## Introduction

Subjective well-being (SWB) is defined as a multidimensional construct, consisted of three major components – satisfaction with life (LS), positive affect (PA), and negative affect (NA) that represent an ongoing state of psychological wellness (Diener, 1984). More recently, satisfaction in specific life domains (henceforth domain satisfaction (DS), e.g., satisfaction with health) was included in the definition of SWB (Diener, Suh, Lucas, & Smith, 1999). Life satisfaction and domain satisfaction are considered cognitive components because they are based on evaluative beliefs (attitudes) about one's life, positive affect and negative affect assess the affective component of SWB (Diener, 1984; Diener, Suh, Lucas, & Smith, 1999).

There are two main theoretical traditions that have contributed to the understanding of SWB (e.g., Diener, & Lucas, 2000). A 'bottom up' perspective mainly analyzed the impact of contextual factors in the SWB of individuals that was explained by a linear combination of domain-specific satisfaction (DS) variables, such as satisfaction with income, housing, or social contacts. Diener (1984) was the first who suggested that the effects could just as well be reversed, that is, go from SWB to DS. He called his model the "top-down" model in contrast to the "bottom-up" model. This suggestion was based on the idea that satisfaction might be determined more by personality characteristics, such as temperament, social comparison, the goal-achievement gap and adaptation, than situational circumstances (see Diener, & Ryan, 2009 for a review). Personality - comprising the psychological aspect of a person that is carried from one situation to another - is one of the strongest and most consistent predictors of subjective well-being (Boyce,

Wood, & Powdthavee, 2013; Diener, & Lucas, 1999; Ferrer-i-Carbonell, & Frijters, 2004; Lykken, & Tellegen, 1996).

The five-factor model of personality (Big Five) is the most widely used taxonomy of personality characteristics (Larsen, & Buss, 2008). It distinguishes five personality traits: neuroticism, extraversion, openness to experience, agreeableness, and conscientiousness. Although the big five model of personality certainly provides results about important personality dimensions, it is just as certain that there are other personality characteristics not embodied by the big five that are just as worthy of study. To date, a great deal has been learned about the personality traits, values, goals, and social behaviors and cognitions of happy individuals compared to their less happy counterparts (e.g., DeNeve, & Cooper, 1998; Diener, Suh, Lucas, & Smith, 1999; Steel, Schnnidt, & Shultz, 2008). Our literature review (Malkina-Pykh, & Pykh, 2013) revealed that the following psychological variables are examined in many studies as predictors of SWB level: locus of control, self-actualization, perfectionism, alexithymia, and sociotropy, and body image dissatisfaction, sense of coherence, emotional intelligence, optimism, self-esteem and self-efficacy. We should note that proposed personality variables are probably best not thought of as a more or less fixed personality trait like extraversion or neuroticism. They are more the dispositions associated with persistence/perseverance, good coping skills, coherent sense of one's personality, and the ability to act in accordance with that personality. Also, it is necessary to stress that overwhelming proportion of the studies on the associations between SWB and personality are conducted in the framework of 'top-down' approach (see Malkina-Pykh, & Pykh, 2013 for review).

Both bottom-up and top-down theoretical approaches, however, ignore the cardinal question of how personality influences SWB. The top-down theory does not explain how and why predispositional reactions to stimuli influence happiness, while the bottom-up theoretical approach does not explain how different people can experience the same events but have very different interpretations. One possible way of better constructing these theories is by linking them to individual mental maps (models). Mental maps are those core beliefs that may explain how individuals select and process information in interpreting life events, and may account for individual differences in these interpretations (e.g. Erez, Johnson, & Judge, 1995).

Recent findings in neuroscience and developmental psychology propose some possible explanations of the sources of these thought processes (mental models). Attachment theory explains how parents and other environmental impact "programmed" the child without conscious discrimination because highly suggestible state of subconsciousness dominated by the emotional brain (right hemisphere) characterizes the child development during the first years of life (Bowlby, 1969; Schore, 2001; 2002). Described as internal working models, these cognitive/affective representations help organize affect and social experience and shape not only current but future interpersonal relationships (Lewis, Feiring, & Rosenthal, 2000; Waters, Merrick, Treboux, Crowell, & Albersheim, 2000). These internal working models are encoded and stored as implicit procedural memories that are unconscious.

The concept of mental maps and implicit memory are also relevant to the model of SWB homeostasis proposed by (Cummins, Gallone, & Lau, 2002). This important theoretical model postulates that SWB fluctuates around a stable set point that is determined by heritable factors such as personality (e.g., Brickman et al. 1978; Headey, & Wearing, 1989; Lykken, & Tellegen, 1996). This integrated model couples a primary genetic capacity (neuroticism, extraversion and positive and negative affect) with a secondary buffering system (the cognitive aspects of control, self-esteem, and optimism). Cognitive buffers influence patterns of thinking and the processing of positive and negative experiences. This assumption about stable set point of SWB are also known as adaptation level theory (Brickman, & Campbell, 1971), dynamic equilibrium theory (Headey, & Wearing, 1989, 1992), and set-point theory (Diener et al. 2006; Lucas, 2007; Lykken, & Tellegen 1996), and became a fundamental part of the SWB framework.

Serious doubts on the validity of set-point theory came up when SWB researchers started analyzing large-scale panel data sets, particularly the German Socio-Economic Panel (SOEP; Headey, 2010; Wagner et al. 2007) and the British Household Panel Study (BHPS; Taylor et al. 2009). Recent research suggests that happiness can, to some degree, be changed (e.g., (Diener, Lucas, & Scollon, 2006). Findings from neuroscience reveal that the brain remains open to new experiences from the environment during the lifespan. This process was called "brain plasticity" and it involves not only the creation of new synaptic connections among neurons but also the growth of new neurons (e.g., Barbas, 1995). These findings have stimulated the development of interventions that aim at increasing SWB at an individual and public policy levels.

Individual and public policy interventions to increase SWB are still a very new field of research, but the first results are promising. Several empirical studies suggest that high SWB has positive consequences: happy people are more likely to get and stay married, to be successful in their jobs, and to earn more money (e.g., Lyubomirsky et al. 2005; Oishi et al. 2007). As psychologists have traditionally operated at an individual level, individual interventions provide tools that each person can apply to become happier (e.g.,

King, 2008; Lyubomirsky, 2008; Lyubomirsky, et al. 2005). A wide variety of interventions have been developed to increase well-being such as techniques based on cognitive behavioral therapy (e.g. Diener, & Ryan, 2009; Huppert, 2005; Layard, 2006), positive psychology interventions (PPIs) (e.g. Seligman, Rashid, & Parks, 2006; Sin, & Lyubomirsky, 2009), well-being (e.g., Fava, & Tomba, 2009) and wellness (Myers, & Sweeney, 2007) therapies.

The interest in body-oriented approaches for increasing subjective well-being has been growing continuously during the past years. Embodiment concept highlights the bidirectional link between the motor and cognitive systems (Koch, & Fuchs, 2011). It means that affect and cognition cause changes in movement, but also movement causes change in cognition and affect via feedback effects. Such movement feedback is defined as the afferent feedback from the body periphery to the central nervous system and plays a causal role in the emotional experience, the formation of attitudes, and behavior regulation. Therefore, there are many reasons to include mind, emotions and body in a successful psychotherapy (Froesch-Baumann, 2002). A great body of literature reveals the importance of engaging in regular physical exercise (aerobic dancing) for improving psychological well-being and overall quality of life (e.g. Behzadnia, Mohammadzade, Farokhi, & Ghasemnejad, 2014). Another promising body-oriented approach for increasing SWB is dance movement therapy (DMT). The findings of the study (Murcia, et al., 2010) revealed that dancing is perceived to be a multidimensional activity that contributes positively to several aspects of human wellbeing. In meta-analysis (Koch, et al., 2014) the effectiveness of dance movement therapy (DMT) and the therapeutic use of dance for the treatment of health-related psychological problems were evaluated and results suggest that DMT and dance were effective for increasing quality of life and decreasing clinical symptoms such as depression and anxiety.

However, there is not much indication in the subjective well-being literature that body awareness, movement, and body-oriented techniques are used in the increasing subjective well-being. Body-oriented interventions are still in the early stages of development, while interventions that address negative affect, depression, and general unhappiness have a long and deep history. herefore, while there are some interventions currently in existence, more interventions are needed (Diener, & Ryan, 2009).

In the present study, we propose the approach of rhythmic movement therapy (RMT) for increasing subjective well-being (Malkina-Pykh, 2012; 2013; 2014). RMT is one of the creative arts therapies based on the 'theoretical interdependence of movement and emotion' (Homann, 2010; Padrao, & Coimbra, 2011) along with dance/movement, music, art and drama. RMT as well as many others body-cantered psychotherapies sees the body as part and parcel of the "mental" processes that govern the flexibility and range of our response patterns (Mowrer, 2008). It means that body literally holds and maintains implicit cognitive, emotional, and perception material that shapes and constrains how we act and that access to and transformation of this material is necessary for increased flexibility and choice in one's life situations (Wylie, 2004).

Detailed description of RMT is provided in our previous studies (Malkina-Pykh, 2001, 2012, 2013, 2014) and here we present only its basic features. RMT is a model of psychological intervention that is philosophically and theoretically rooted in body-oriented psychotherapy (Lowen, 1975; Reich, 1949; etc.), dance movement psychotherapy (Chodorow, 1991; Stark, 1987; etc.), and rhythmic gymnastics (aerobics)' (Malkina-Pykh, 2001). The therapeutic work in RMT includes two main components: (1) diagnostic system of core personal problems corresponding with various characters and body types and (2) rhythmic movement as a medium of change.

Practical applicability of RMT theory and practice is based on the recent advances in developmental psychology and neuroscience such as attachment styles, implicit memory and embodiment mentioned above. RMT character typologies follow developmental models proposed by S. Freud (1962), E. Erikson (1950) and K. Horny (1937). RMT diagnostic system incorporates Freud's oral, anal, and genital stages and three stages of Erikson psychosocial model applied to the developmental period from in birth to about 7 years of age, as well as Kretschmer's (1925) typology and somatotypes proposed by Sheldon (1940) which allows associating preliminary the core personal problems with specific body types (Malkina-Pykh, 2001). Based on the mentioned descriptions of character types, we proposed a combined trifold diagnostic system that allows us to understand the major problems of each of three character types with the specific body types: (1) orals (asthenic/leptosomic, ectomorphic body type, cerbrotonia) – have problems with dependency needs (boundaries, communication), (2) anals (athletic, mesomorphic body type, somatotonia) – have problems with aggression and overcontrol and (3) phallics (pyknic, endomorphic body type, viscerotonia) – have problems with ego-identity (self-image). Also, other three core problems are common for all character types: grounding, blockage release and integration (Malkina-Pykh, 2001). In more details RMT diagnostic system is described in our previous study (Malkina-Pykh, 2013).

RMT proposes theory, methods, and techniques that allow individuals access deep implicit material and transform related internalized emotional schema (core beliefs). The basic mechanisms of RMT are the rhythmic movement itself and kinesthetic trance. The focus of the RMT is to make core beliefs and emotional memories explicitly conscious where they can be transformed into more flexible and functional understandings. The intent is to help the client to establish and stabilize new beliefs, thereby diminishing and ending the habituated power of old beliefs and implicit emotional material.

Recent study (Malkina-Pykh, 2012) on the application of RMT in weight reduction program revealed that improvements achieved in RMT group was not restricted to the disordered eating symptomatology and obesity, but was found also in body image dissatisfaction, alexithymia and four dimensions of perfectionism. The findings of another study (Malkina-Pykh, 2013) demonstrated a significant decrease of alexithymia and changes in associated personality variables as a result of RMT intervention.

In addition to theory and methods of psychotherapy one of the most important theme in psychotherapy researches is the assessment of treatment outcome. At the moment mainly three groups of statistics are used to express the consequences of psychological interventions for participants: statistical significance of within- and between-group differences, effect size (ES), and clinical significance (CS). Although statistical significance tests and ES indices provide important information regarding mean differences, they do not give information regarding the variety of responses to treatment within the treated group. Thus, these types of aggregate data analysis are not useful for determining individual changes, which can be very valuable in clinical practice for treatment planning, monitoring course of illness and evaluating response to treatment (Lambert, & Ogles, 2009). Clinically significant change refers to changes in patient functioning that are meaningful for individuals who undergo psychosocial or medical interventions. In this regard, it allows researchers to focus on the functioning of each patient rather than on group averages and statistical significance of between-group comparisons (Bauer, Lambert, & Nielsen, 2004; Eisen, Ranganathan, Seal, & Spiro, III, 2007).

Several criteria for clinically significant change exist (see Bauer, Lambert, & Nielsen, 2004; Wise, 2004 for reviews). The most popular method for estimating clinically significant change was proposed by (Jacobson, Follette, & Revenstorf, 1984; Jacobson, & Truax, 1991) and suggested a two-step criterion for clinically significant change. First, a cut-off point for a measure of psychological functioning is established that is conceptualized as a cut-off between two populations: a patient/dysfunctional population, and a nonpatient/functional population. The second step of the Jacobson–Truax (JT; 1991) method is to determine whether a client's change from pre-test to post-test is reliable rather than simply an artifact of measurement error. To assess this, Jacobson et al. (1984) proposed a reliable change index (RCI) that each participant has to pass to demonstrate that his or her change is not simply due to chance. RCI is calculated as follows:

$RCI = \dfrac{x_2 - x_1}{S_{diff}}; \; S_{diff} = \sqrt{2(SEM)^2}; \; SEM = s_x\sqrt{(1-r_{xx})},$ where $x_1$ and $x_2$ are pre-treatment and post treatment scores, $SEM$ is the standard error of measurement, $s_x$ is the standard deviation of pre-treatment group, $r_{xx}$ is the internal consistency of the test (Cronbach's $\alpha$) and $S_{diff}$ is a variation of the standard error of measurement. If the absolute value of RCI is greater than 1.96, then change is considered statistically reliable. Based on the two criteria (cut-off and RCI), the JT method classifies individuals as *Recovered* (i.e., passed both cut-off and RCI criteria), *Improved* (i.e., passed RCI criterion but not the cut-off), *Unchanged* (i.e., passed neither criteria), or *Deteriorated* (i.e., passed RCI criterion but worsened).

Although the methods for estimating clinically significant change could be regarded as a step further comparing with statistical significance in assessing the results of the treatment, all assessment methods mentioned above, even clinical significance, are prospective or retrospective, but not predictive in any other form than probabilities and does not yield information about linkages between causes and effects, especially in case of nonlinearity of interactions within system under study. The limitations of such models as exploratory and predictive tools are well known and describe elsewhere (e.g., Seber, & Wild, 2003).

The objective of the present study was to apply the nonlinear response function model of subjective well-being (RFSWB model) to evaluate the outcome of rhythmic movement therapy (RMT) for increasing subjective well-being and to analyze whether intervention-related changes in several psychological variables were mechanisms underlying SWB increase in subjects participating in RMT group.

## Response Function Model of Subjective Well-Being (RFSWB model)

The procedure of RFSWB model development was described in details in our previous studies (e.g. Malkina-Pykh & Pykh, 2013). Our operationalisation of SWB determines it as a multidimensional variable that is composed of evaluations about different domains of life satisfaction in a bottom-up or component-based approach, where participants appraise in a cognitive and affective way how they experience their lives. Our operationalisation of SWB includes five domains: material, health, work, leisure and recreation and personal competence. Furthermore, our operationalisation of overall SWB (evaluated in domains) includes dimensions of several psychological constructs that are hypothesised to be components of the mental map underlying SWB, namely the level of selfactualisation, sociotropy, perfectionism, locus of control, body image dissatisfaction, neuroticism and alexithymia (Malkina-Pykh, & Pykh, 2013).

RFSWB model is based on method of response function (MRF), flexible nonlinear regression method which allows prediction of the impact of the subjects' psychological variables as well as their changes on the actual level of SWB. Thus MRF may be used to identify and characterize the effect of potential prognostic factors on treatment outcome variable. Generally speaking, MRF is a method of the construction of purposeful, credible integrated models from data and prior knowledge or information. Integration means capturing as much as possible of cause-effect relationships and describing them with an operator of transition, or ''input–output'' function. Data series observations contain "hidden" information on the processes under consideration and one of the main purpose of the proposed method is to ''extract'' and describe these hidden relationships. The MRF models nonlinear relationships among variables, can handle nominal or ordinal data, and does not require multivariate normality. This approach allows us to take into account main essential features of psychological systems: complexity, multidimensionality, uncertainty, irreducibility, and so on. Also we argue that MRF fills the gap between linear statistical techniques and full complexity models such as psychophysical models, neural modeling, etc.

Let us assume the basic definitions of the MRF. Factors are system's properties that directly affect processes or characteristics under study. The factors are designate as a vector $x = (x_1, x_2, \ldots, x_n)$. Then, a function which depends on a single active factor, i.e. the function of a single variable $f_i(x_i)$ is defined as a ***partial response function*** of the characteristic or the process under study. A function $F(x_1, \ldots, x_n)$ which accounts for all the factors considered and presented as a combination of partial response functions $f_i(x_i)$ is defined as ***generalized response function.***

Now the generalized response function is proposed in the form

$$F(x_1, \ldots, x_n) = \prod_{i=1}^{n} f_i(\alpha^i, x_i), \qquad (1)$$

where $n$ is the number of the factors under study, $\alpha^i$ is a vector of parameters, the values of which we have to determine in the process of identification. Basically, it has been criticized that the multiplicative form represents the independence of the influencing factors. Previous studies demonstrated that this problem could be resolved successfully using some specific technique for the evaluation of parameters of the generalized response function $F(x_1, \ldots, x_n)$ (Malkina-Pykh & Pykh, 2013). The additional restriction in the identification procedure was introduced:

$$\max_{x_i} f_i(\alpha^i, x_i) = 1 \qquad (2)$$

It is evident that standardization condition (2) gives us a possibility to compare the impact of different factors on the process under study. The theory of the method of response function and its applications in details has been described in several articles and monograph (e.g. Malkina-Pykh & Pykh, 2013).

Based on our operationalisation of SWB the RFSWB model is looking as follows:

$$IISW_{mod} = IISW_{max} \cdot F_{st}$$

$$F_{st} = f_1(I) \cdot f_2(Te) \cdot f_3(S) \cdot f_4(PS) \cdot f_5(MD) \cdot f_6(LC) \cdot f_7(B) \cdot f_8(A) \cdot f_9(N)$$

$$f_j(x_j) = \alpha_j \cdot x_j^{\beta_j} \cdot \exp\left\{-c_j\left\{\frac{x_j^{\gamma_j}}{x_{j\max} - x_j}\right\}\right\} \quad j=1,2,6$$

$$f_5(MD) = \alpha_5 \cdot MD^{\beta_5} \cdot \exp\left\{-c_5\left\{\frac{MD}{MD_{\max} - MD}\right\}^{\gamma_5}\right\}$$

$$f_j(x_j) = \alpha_j \cdot (b_j - c_j \cdot (1 - \exp(-d_j \cdot x_j^{\gamma_j}))) \quad j=3,4,9$$

$$f_j(x_j) = \alpha_j \cdot (b_j - c_j \cdot (1 - \exp(-d_j \cdot x_j))^{\gamma_j}) \quad j=7,8 \qquad (3)$$

where $IISW_{mod}$, $IISW_{max}$ are the modeled actual values of SWB and maximal possible value obtained during the estimation of model's parameters respectively, $I$ are the scores of inner-directed scale, $Te$ are the scores of time competence scale, $S$ are the scores of sociotropy scale, $PS$ are the scores of personal standards scale, $MD$ are the scores of concern over mistakes, doubts about actions scale, $LC$ are the scores of general locus of control scale, $B$ are the scores of body image dissatisfaction scale, $A$ are the scores of alexithymia scale, $N$ are the scores of neuroticism scale, $F_{st}$, $f_j$ are the generalized and partial response functions respectively, $\alpha_j, b_j, c_j, d_j, \gamma_j, x_{j\max}$ are parameters for evaluation, $j = 1,..,9$.

Estimation of model parameters and its validation have been provided in our previous study (Malkina-Pykh & Pykh, 2013). The goodness of model fit (i.e., average value of residual errors) is 6.6% (range from 0.35% to 17.3%) between model and data. While the standard error of SWB measurement in the given sample is equal to 6.9%, the goodness of RFSWB model fitting seems satisfactory. Table 1 contains the values of RFSWB model parameters. The estimated value of $IISW_{max}$ is equal to 18.4. The average error of validation of RFSWB model is 10.5% (range from 0.3 to 23.5%) between model and data. The comparison seems satisfactory.

**Table 1 Values of SWB model parameters**

| Parameter | Values | Parameter | Values |
|---|---|---|---|
| $\alpha_1$ | 0.20 | $c_5$ | 0.06 |
| $b_1$ | 0.86 | $\gamma_5$ | 2.87 |
| $c_1$ | 0.18 | $MD_{max}$ | 15.3 |
| $\gamma_1$ | 0.79 | $\alpha_6$ | 0.63 |
| $x_{1\max}$ | 12.6 | $b_6$ | 0.33 |
| $\alpha_2$ | 0.97 | $c_6$ | 0.09 |
| $b_2$ | 0.04 | $\gamma_6$ | 1.29 |
| $c_2$ | 0.05 | $x_{6\max}$ | 12.8 |
| $\gamma_2$ | 0.86 | $\alpha_7$ | 4.05 |
| $x_{2\max}$ | 12.6 | $b_7$ | 0.25 |
| $\alpha_3$ | 0.33 | $c_7$ | 1.66 |
| $b_3$ | 2.99 | $d_7$ | 0.13 |
| $c_3$ | 0.73 | $\gamma_7$ | 7.59 |
| $d_3$ | 0.09 | $\alpha_8$ | 0.83 |
| $\gamma_3$ | 3.21 | $b_8$ | 1.20 |
| $\alpha_4$ | 0.44 | $c_8$ | 1.80 |
| $b_4$ | 2.26 | $d_8$ | 0.26 |
| $c_4$ | 0.55 | $\gamma_8$ | 12.4 |
| $d_4$ | 0.28 | $\alpha_9$ | 0.92 |
| $\gamma_4$ | 1.07 | $b_9$ | 0.93 |
| $\alpha_5$ | 0.95 | $c_9$ | −0.15 |
| $b_5$ | 0.03 | $d_9$ | 0.0000001 |
| | | $\gamma_9$ | 7.54 |

Figure 1 presents the graphs of partial response functions of the personality variables under consideration.

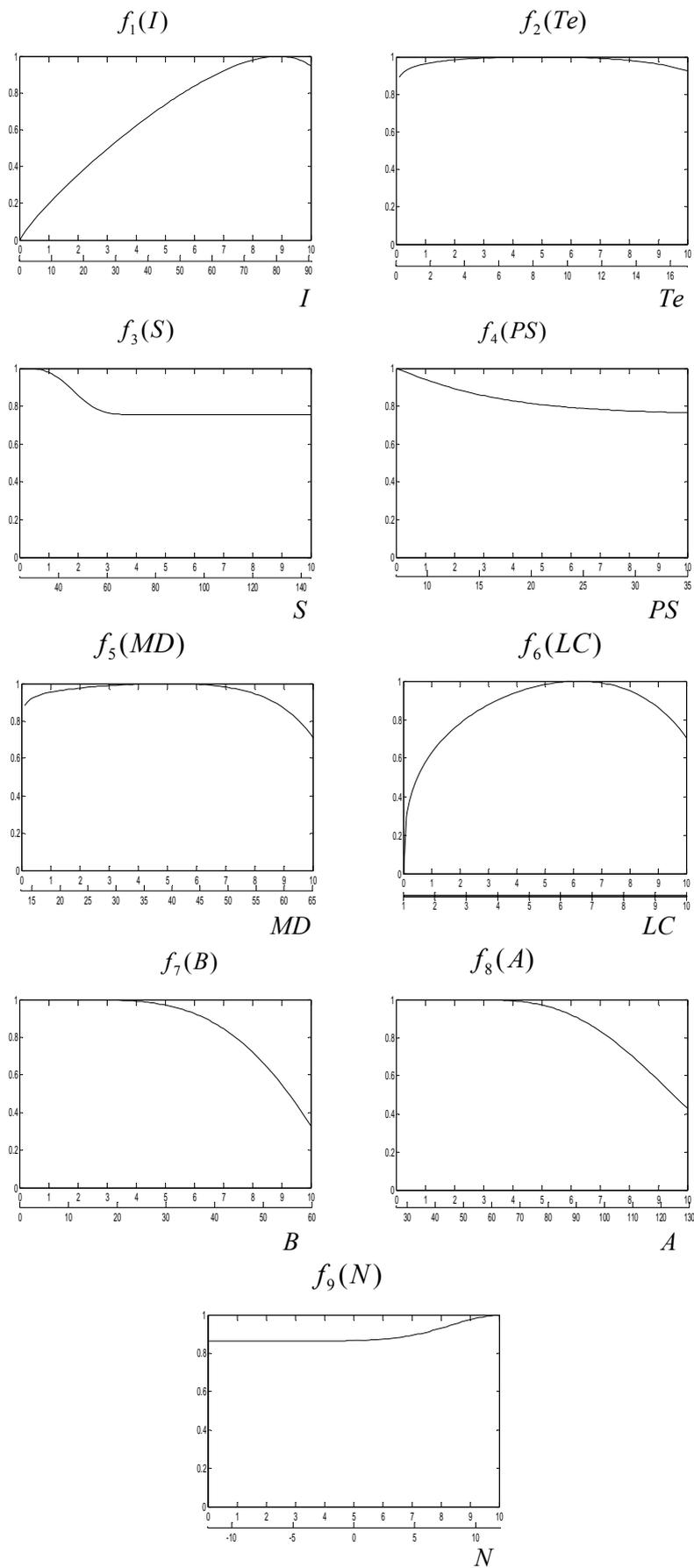

**Figure 1** Partial response functions of the RFSWB model. *I* are the scores of inner-directed scale, *Te* are the scores of time competence scale, *S* are the scores of sociotropy scale, *PS* are the scores of personal standards scale, *MD* are the scores of concern over mistakes, doubts about actions scale, *LC* are the scores of general locus of control scale, *B* are the scores of body image

dissatisfaction scale, *A* are the scores of alexithymia scale, *N* are the scores of neuroticism scale, $f_j$ are the partial response functions respectively, $j = 1…9$.

The obtained views of partial repsonse functions will be described in details in Discussion section.

# This study

**Method**

Participants

The study was conducted in the framework of Mental Health Management Program organised in the "Human Ecology" department of the Research Center for Interdisciplinary Environmental Cooperation of Russian Academy of Sciences, St-Petersburg, Russia.

At baseline, a total of 273 subjects were recruited among those searching for counseling (psychotherapy) regarding various non-clinical psychological problems: e.g. low self-esteem, family problems, workplace bullying, etc. No any medical disorders or clinical complaints (e.g. diabetes mellitus, rheumatoid arthritis, pulmonary disease, depression, panic attack, etc.) were claimed by the participants. Demographic items included age and gender, which were assessed with single questions. The participants were aged between 20 and 60 years old, mean age was 37.3±10.5 years, 54 (20%) were males and 219 (80%) were females.

The study was approved by the Ethics Committee of I.I. Mechnikov North-West State Medical University, St.-Petersburg, Russia, and was performed in accordance with the ethical standards laid down in the 1964 Declaration of Helsinki. All participants signed informed consent form before participating in the study.

**Measures**

Subjects were assessed with measures listed below. Russian-validated translations of all measures were used.

SWB was measured using *Integral Index of Social Well-being* (IISW) (Panina, & Golovakha 2001). The test includes 20 items based on a three-point Likert scale and covers five domains of SWB: work, material well-being, health, leisure/recreation and personal competence. Examples of IISW items include 'How satisfied are you with your job?', 'How satisfied are you with your health?', 'How satisfied are you with the amount of leisure in your life?', 'How satisfied are you with your self-confidence?' etc. The responses are tabulated as follows: 1= 'not satisfied'; 2= 'don't know (not of interest)'; 3= 'satisfied'. Higher scores show higher level of SWB, maximum score is 60.0. The IISW has demonstrated strong internal consistency and test-retest reliability (0.67). In the experimental sample, the IISW items generated alpha coefficients of 0.87 (Panina, & Golovakha 2001).

The *Personal Orientation Inventory* (POI) (Shostrom, 1974). In the present study, only the main scales were used because Shostrom (1974) recommended that the Time Competence (*Tc*) and Inner-Directed (*I*) scales may be used when a quick estimate of examinees' levels of self-actualization is desired.

General Locus of Control Scale *of the Locus of Control Inventory* (LOC) was used to measure locus of control because it uses all the items in the LOC and is the only LOC scale that does not have overlapping items (Rean, 2001).

The Neuroticism (N) Scale was selected from the *Eysenck Personality Inventory* (EPI; Eysenck, & Eysenck 1963) to measure the traits of extraversion–introversion and neuroticism. Scores range from -12 to 12.

Alexithymia was investigated using the *Toronto Alexithymia Scale-26* (TAS-26) (Taylor, Ryan, & Bagby, 1985). The TAS-26 is a 26-item self-report measure of alexithymia with a three-factor structure theoretically congruent with the alexithymia construct. A cut-off score of 62 was used to define alexithymia as recommended.

*Body Image Test* (Jade, 2002) was used to investigate body image dissatisfaction. Participants answered how often they felt uncomfortable about their appearance in different situations. The test includes 20items, which are answered based on a four-point Likert scale. Higher scores show greater body image dissatisfaction.

*Personal Perfectionism Scale* (PPS) includes three subscales from the Multidimensional Perfectionism Scale by Frost, Marten, Lahart, & Rosenblate (1990): 'Personal Standards', 'Concern over Mistakes' and 'Doubts about Action'. 'Concern over Mistakes' and 'Doubts about Actions' subscales were combined into one scale. Thus, PPS includes 20 items, which are answered based on a seven-point Likert scale. The reliability and validity of the Russian version of the PPS were described in our previous study (Malkina-Pykh, 2012).

Sociotropy Scale of *Personal Style Inventory* (PSI; Robins, Ladd, Welkowitz, Blaney, Diaz, & Kutcher, 1994) was used to assess the constructs of sociotropy. The PSI Sociotropy Scale consists of 24 items which are rated on a six-point Likert scale. Higher scores show greater chance of sociotropy.

Simple *Symbol Personality Test* was applied for express diagnosis of the subject's character type in addition to body type diagnostics. *Symbol Personality Test* is the modification of *Symbol Test* (Funch, 1995) and includes the pictures of three geometric figures: triangle, square and circle. We have the person look briefly at these three figures and ask him to choose the one he likes best. Triangle is correspondent with oral, square with anal and circle with phallic character type.

Participants were given approximately 1 hour to complete the scales described above.

**Design**

After baseline assessment all subjects of experimental group were divided into two more groups in accordance with their SWB level: group 1 – (20-40 IISW scores mean very low, low and medium SWB levels) – 156 subjects, 120 females (77%), 36 males (23%), mean age 37,6±11,8 years) and group 2 – (41-60 scores of IISW mean high and very high SWB levels) – 117 subjects, 99 females (85%), 18 males (15%), mean age 36,8±8,54 years.

After that RMT program was proposed to the subjects of the first group for increasing their SWB level. Participant inclusion criterion included very low, low and medium SWB scores on IISW. Exclusion criterion included any medical disorders or clinical complaints (e.g. diabetes mellitus, rheumatoid arthritis, pulmonary disease, depression, panic attack, etc.). All 156 subjects were willing to participate and were randomly assigned ten RMT intervention groups (105 subjects, 24 males, 81 females, mean age 37.6±11.7 years) and five waiting list groups (51 subjects, 12 males, 39 females, mean age 37.8±11.9 years). Waiting list groups didn't receive any counseling or treatment and were contacted after 4 months by psychologists only to administer the assessment scales. The waiting list groups received RMT intervention after completion of post-assessment. Subjects with high levels of SWB (N=117) were administrated to standard cognitive-behavioural counseling.

Experimental group in total included 12 subjects with oral character type, 78 subjects with anal character type and 66 subjects with phallic character type. One-way ANOVA with character type as a factor was applied to analyse the distribution of low levels of SWB. Low levels of SWB are equally presented in all character type groups: orals – 39.5±.52, anals – 38.9±3.6, phallics – 37.7±4.3 scores on IISW ($F$=2.26, $p$=0.11). Nevertheless, we argue that the underlying mechanisms of low SWB are correspondent with the core problems of each character type.

The RMT intervention was delivered by psychologists who were trained in RMT during a 1-year course. Following the proposed diagnosis system, the main objects of work in RMT are core problems of boundaries and communication for orals, for annals – core problems of aggression and overcontrol (for annals), and for phallics – core problem of self-image (for phallics). Also RMT works with three core problems common for all character types: grounding, blockage release, and integration.

The present RMT intervention consists of 16 once-a-week structured sessions of 45-50 min each in a group setting. Each session follows the same basic structure and is devoted to one selected core problem. In the present intervention 2 sessions were devoted to each core problem. They start with the introduction into the theory of the problem under consideration given by the psychologist (first session on each core problem) and/or with the verbal exchange and feed-back in a circle.

Each RMT session consists of three types of exercises: diagnostic, resource, and test. As an example, we describe sessions devoted to communication patterns of participants.

The main object of *diagnostic exercises* in the present context is the discovering the usual patterns of subject's communication and their corresponding connection to personality type that allows the emergence of a potential framework for the course of therapy. One possible option for communication diagnostic exercise is sirtaki dance. Diagnostic exercise in this session allows for greater awareness of how the patient normally communicates.

*Resource exercise* in this session includes highly structured steps and choreography of kick rock'n-roll dance. One of the most important aspects of kick rock'n-roll is the demand of high level of communication skills in participants. By utilizing the rhythmic movement in *resource exercises*, patients increase their communication skills that lead to better understanding and expression of feelings, develop coping skills, increase focus and concentration, and accept limitations.

*Test exercises* are intended to analyze whether the increase in patients' communication skills is producing a significant improvement in their touch with reality, interpersonal relationships, in the quality of their thinking and feeling, and in their subjective well-being.

The movement part closes with verbal feed-back which lasts about 10–15 minutes. The aim of this part is to exchange the movement experience by participants in a circle in order to promote as well as to integrate new interactive experience. This phase is devoted to reflection and discussion about whatever events had occurred in the session. There are discussions regarding the conflicts appearing during the activities and ways to resolve them, the interactions involved, the degree of pleasure induced by the exercises, and so on.

All patients of RMT group completed the intervention programs in accord with preplanned schedule of weekly 16 sessions. After this stage was completed, patients of RMT and control groups were repeatedly assessed with the same measures as at the beginning of the Stage I of the experiment.

## Analysis

To support our hypothesis that the selected psychological variables could be the element of mental map responsible for evaluation of subjective well-being we applied one-way ANOVA with a group as a factor (high SWB group vs. low SWB group) to get the evidence. However, the main assessment method in evaluating the outcome of RMT for subjective well-being was the application of SWB model in the RMT intervention groups. Computer experiments with RFSWB model and pre scores of psychological variables of participants of RMT groups were provided with the aim to reveal how satisfactory RFSWB model predicted the initial level of subjective well-being. Computer experiments with RFSWB model and post scores of psychological variables of participants of RMT groups were provided with the aim to demonstrate whether intervention-related changes in several psychological variables were mechanisms underlying SWB increasing in subjects participating in RMT program. JT method was used in our study as the criterion for clinically significant change to analyze weather RFSWB model simulates it satisfactorily. The standard error of measurement (*SEM*) in the given sample is equal to 1.7 and Cronbach's $\alpha$ is equal to .83 (Panin, & Golovaha, 2001).

## Results

There was no drop-outs of participants during the study. After the groups 1 and 2 were formed, the results of one-way ANOVAs with group as a factor in entire sample indicated no differences in SWB level due to age ($F(1,271)=0.52$, $p=0.47$) or gender ($F(1,271)=0.05$, $p=0.82$). The ANOVAs demonstrated that subjects of group 1 compared with the subjects of group 2 have significantly different levels of SWB as well as of all other personality variables, except for "Personal standards" (Table 2). Thus, we suppose that our specific operationalisation of SWB construct has merit.

**Table 2. Assessment scores in groups 1 and 2**

| | Psychological variables and IISW | Group 1 N=156 M (SD) | Group 2 N=117 M (SD) | ANOVA *F(1,271)* | *P* |
|---|---|---|---|---|---|
| 1. | Integral index of social well-being | 36.5(4.3) | 48.7(5.1) | 459.7 | <0.001 |
| 2. | Time competence | 6.1(2.31) | 9.1(2.83) | 95.9 | <0.001 |
| 3. | Inner-directedness | 37.0(2.26) | 48.2(9.87) | 145.8 | <0.001 |
| 4. | Sociotropy | 96.7(13.2) | 82.8(16.9) | 53.9 | <0.001 |
| 5. | Personal standards | 22.1(4.49) | 21.0(5.07) | 3.27 | 0.071 |
| 6. | Concern over mistakes/doubts about actions | 43.8(7.39) | 29.9(9.14) | 192.0 | <0.001 |
| 7. | General locus of control | 3.77(1.19) | 6.0(1.56) | 179.7 | <0.001 |
| 8. | Body image dissatisfaction | 25.4(10.2) | 15.8(11.7) | 51.7 | <0.001 |
| 9. | Neuroticism | 4.92(4.75) | −0.13(4.29) | 82.1 | <0.001 |
| 10. | Alexithymia | 72.8(11.0) | 58.5(11.6) | 107.3 | <0.001 |

The average error (i.e., average value of residual errors) of RFSWB model simulation of subjective well-being before RMT intervention is equal to 9.4% (range from 0% to 30.0%) between model output and experimental data. The average error of RFSWB model simulation of subjective well-being after RMT intervention is equal to 1.3% (range from 0% to 23.7%) between model output and experimental data. While the goodness of RFSWB model fit is equal to 6.6% (i.e. average value of residual errors in estimation of model's parameters) (Malkina-Pykh & Pykh, 2013) the goodness of RFSWB pre-post simulations was accepted as satisfactory. The summary results of the application of JT method to SWB experimental scores and SWB modeled scores are presented in Table 3.

**Table 3 Percentages and frequencies of clinical significance classification for Jacobson–Truax method for experimental IISW scores and SWB model scores**

| Group | Scores | Approach | Deteriorated | | Unchanged | | Improved | | Recovered | |
|---|---|---|---|---|---|---|---|---|---|---|
| | | | % | Freq. | % | Freq. | % | Freq. | % | Freq. |
| RMT (N=105) | IISW exp | JT | 0 | 0 | 0 | 0 | 4 | 4 | 96 | 101 |
| | RFSWB model | JT | 0 | 0 | 0 | 0 | 7 | 7 | 93 | 98 |
| Control (N=51) | IISW exp | JT | 0 | 0 | 94 | 48 | 3 | 3 | 0 | 0 |
| | RFSWB model | JT | 0 | 0 | 92 | 47 | 4 | 4 | 0 | 0 |

Note: IISWexp are the experimental scores of IISW scale, RFSWB model are the modeled scores; JT = Jacobson–Truax method.

Application of JT to IISW scores in RMT group revealed 0% of deteriorated, 0% of unchanged, 4% of improved and 96% of recovered clients. Application of JT to RFSWB model scores revealed very similar results, such as 0% of deteriorated, 0% of unchanged, 7% of improved and 93% of recovered clients. Also, it is necessary to mention that the results revealed 3 clients who passed the cutoff of IISW scale, but not RCI threshold and 4 such clients for RFSWB model results. Application of JT to IISW scores in control group revealed 0% of deteriorated, 94% of unchanged, 3% of improved and 0% of recovered clients. Application of JT to RFSWB model scores revealed very similar results, such as 0% of deteriorated, 92% of unchanged, 4% of improved and 0% of recovered clients.

## Discussion

The objective of the study was to evaluate the outcome of rhythmic movement therapy (RMT) for increasing subjective well-being using nonlinear regression model of subjective well-being (RFSWB model) as an assessment tool.

The nonlinear modeling procedure described here is useful for several reasons. First, it provides information about the nonlinear relationships between prognostic factors and subjective well-being that is not revealed by the use of standard statistic and/or linear modeling techniques. Second, RFSWB model allows predict whether intervention-related changes in several psychological variables are mechanisms underlying subjective well-being change in subjects participating in RMT intervention program. Third, application of the RFSWB model as assessment tool allows identification of reliable predictors of success in RMT intervention that could help future programs in several ways. For example, an increased focus could be placed on those intervention components more likely to produce desired outcomes while discarding redundant ones. With a description of critical mechanisms of change, researchers and counselors may not only improve treatment efficacy and cost-effectiveness but also gain insight into aspects that contribute to the persistence of the problem in the first place.

The views of partial response functions obtained as a result of model parameters' identification are corresponding with underlying psychological theories except for variable "neuroticism".

The unimodal function was selected to describe the impact of inner-directedness on SWB, because of the idea of pseudo-self-actualization proposed by Shostrom (1974), for those subjects who obtained very high scores on the scale.

Also, the unimodal function was selected for time competence impact on SWB because of the same considerations as for inner-directedness. But in contrast to inner-directedness the results of parameters' identification procedure for time competence partial response function found nearly non-existent impact on SWB level.

One of the possible explanations of the current finding is the inner contradictions of the time competence phenomena. One of the unanswered questions with regard to time perspective concerns the relationship between different temporal orientation profiles with well-being (see Boniwell et al. 2010 for a review). The literature is divided on whether it is the future, present, or the past orientation that is most conducive to well-being. In sum, our findings with regard to the present time perspective are also inconclusive and would benefit from further investigation.

The partial response of sociotropy decreases with increasing sociotropy level until the score of 60 points and then remains stable. In general, this result corresponds with the theoretical considerations that imply the normal range of sociotropy at 72–93 points of the scale. Our findings are consistent with the other results (e.g., Beck, et al. 1983; McBride, et al. 2005; Sato, & McCann, 1998).

Our findings of the partial response function of self-oriented perfectionism (personal standards) are consistent with other data showing a positive association between this variable, psychopathology, and depression (Hewitt, & Flett, 1991; Hewitt, et al. 1996). Self-oriented perfectionism has also been shown to interact specifically with achievement-related stressors to predict severity of depression (Hewitt, & Flett, 1993).

The dimension of perfectionism, namely "concern over mistakes, doubts about action", does not show any significant impact on the level of SWB. It was shown that unique profiles of perfectionism exist within individuals, but that the relative healthiness of personal standards varies as a function of the other dimensions of perfectionism included, such that personal standards did not appear to be associated with poorer health or well-being unless it was combined with high levels of socially prescribed perfectionism.

The partial response function of body image dissatisfaction is consistent with theoretical considerations. In accordance with the results of body image dissatisfaction, the subjects who scored less than 20 points are living comfortably in their body most of the time. Starting from 20 points, the body image dissatisfaction is increasing and SWB is decreasing correspondently. The score of >40 points indicates a condition called body dysmorphic disorder (BDD), which is translated as body hatred.

Our finding on locus of control impact on SWB is consistent with other results, revealing that internals have significantly better SWB than externals. People high on internal locus of control are more active in attempting to manipulate their environment, while externals are passive in manipulating their environment (Phillips, 1980; Reker, 1977; Sammon, et al. 1985; Yarnell, 1971).

Alexithymia partial response function reflects the theoretical cut-off score of 62 used to define between alexithymic and nonalexithymic types. After this point, the response function decreases with increasing alexithymia level, and the SWB level decreases correspondingly. Our findings are generally consistent with previous studies demonstrating that quality of life is related to the emotional as well as to the cognitive dimensions of alexithymia (Henry, et al., 2006). Individuals with alexithymia reportedly used more negative strategies to modulate affect and were also less satisfied with life.

The most unexpected results were obtained on the partial response function of neuroticism and its impact on SWB level. In contrast with the existing dominant theories of the impact of neuroticism on SWB and the initial form of the function (generalized Weibull function), our results showed a small but definite increase in the SWB level with an increase in neuroticism.

However, an analysis of the literature showed that our findings are not unique. The recent study on the revision of set-point theory of well-being by Headey (2007, 2008a, b) showed some kind of similar results. The hypothesis of these studies suggested that if high O is combined with high N, this further increases the "downside risk" of adverse events and substantial decline in long-term SWB. So, in testing the hypotheses, the author included a multiplicative term (N*O) in the analysis. The proposed hypothesis was not supported. The sign of the interaction term was contrary to hypothesis: the sign for N*O was expected to be negative but turned out positive. The author concluded that results relating to interaction terms were problematic, in part due to the danger of multicollinearity.,Our counter-intuitive finding on neuroticism supports the idea that in case of interactive relationships of several personality traits or dispositions the results may differ significantly from the case when the factors are analyzed independently.

Obviously, the results on time competence, maladaptive perfectionism, and neuroticism and their impact on SWB require additional experimental and theoretical studies.

Substantial changes in SWB and underlying psychological state were observed among the participants as a result of RMT intervention. In particular, inner-directedness, time competence and locus of control improved, and sociotropy, perfectionism dimensions, body dissatisfaction, alexithymia and neuroticism decreased. The inclusion of a measurement of such an intrapsychic theoretical construct is of special significance. These findings may support the assumption that solving underlying problems can reduce overt behavioural symptoms, even if the latter are not directly focused on in the session.

Body-oriented psychotherapies for increasing SWB are rather rare. Nevertheless, our findings of RMT effectiveness are generally consistent with the results of several other studies in the field. RMT relies on body-based mindfulness as a primary tool to explore the implicit beliefs that organize life experiences and to address the attachment injuries that shape our emotional realities. Mirror neuron research has pointed toward a strong neuronal connection between one's own motor experience and intersubjective and empathic processes (Gallese, 2003). Coordinated rhythmic movement is the strongest behavior to unite humans, because humans have the capacity to become entrained with one another or with an external stimulus. Entrainment provides a mechanism for physical mirroring, as in gestural mimicking in communication or in dance, and for metaphorical mirroring, as in empathy (Gallese, Keysers, & Rizzolatti, 2004). The findings of the present study support our previous results on the application of RMT for the treatment of disordered eating behaviors and obesity (Malkina-Pykh, 2012) and alexithymia (Malkina-Pykh, 2013).

Our results reveal that RFSWB model predicts satisfactorily not only actual levels of SWB, but clinically significant change as well. It was argued in several studies that the rates of reliable or clinically significant change of a particular sample should not be calculated by summing up the results of the individual participants of that group. This would result in an underestimation of the true rates of change (Bauer, Lambert, & Nielsen, 2004). Estimates of clinically significant improvement for groups of patients affect the degree to which treatments are generally considered to be effective or in need of modification. Timely knowledge about progress and treatment outcomes for individuals can provide opportunities for clinicians to improve care for individuals, utilize individual outcome data to guide future treatment, and engage consumers in the treatment process (Eisen, Ranganathan, Seal, & Spiro, III, 2007). RFSWB model allows to conduct a dose-response study, rather than an "all or none" (intervention vs. control) comparison commonly used in clinical trails. In our previous study (Malkina-Pykh, 2014) we applied the traditional statistical analysis for within- and between groups comparisons (ANOVA and ANCOVA) for evaluation of the RMT effectiveness for increasing SWB in the same sample. Although the results were more impressive in total, they didn't allow estimating the individual results of each subject.

## Study limitations and future research directions

The study limitations highlight the need for future research in this area of SWB. One limitation of our research is that we only provide a partial explanation for the influence of thought processes (personality) on SWB. In general, individual differences play an important role in determining the manner in which people react on life circumstances and then play out in turn in the SWB that is experienced. Unfortunately, there is no consensus upon definition of personality but rather there are almost as many definitions as there are personality theorists and researchers (Pervin, 1989). While other thought processes (dispositions) may also influence the subjective well-being, this must be an area for future research.

Second, IISW doesn't include family domain, although many SWB studies argue that family is one of the most important domain. Third, the study sample is small, limiting interpretation of comparative results and generalization of study findings.

The conclusions regarding the efficacy of the RMT evaluated in this trial are limited by the absence of follow-up data. Though we are currently gathering follow-up data on the proposed treatment, longer studies with follow-ups are needed to better assess the overall effectiveness of the RMT approach.

Malkina-Pykh, I.G. (2014). Effectiveness of rhythmic movement therapy: Case study of subjective well-being. *Body Movement and Dance in Psychotherapy*, http://dx.doi.org/10.1080/17432979.2014.977822.

McBride, C., Bacchiochi, J.R., & Bagby, R.M. (2005). Gender differences in the manifestation of sociotropy and autonomy personality traits. *Personality and Individual Differences*, *38*, 129–136.

Mowrer, J. (2008). Accessing implicit material through body sensations: The body tension sequence. *Hakomi Forum*, Issue 19-20-21.

Murcia, C.Q., Kreutz, G., Clift, S., Bongard, S. (2010). Shall we dance? An exploration of the perceived benefits of dancing on well-being. *Arts & Health, 2*(2), 149–163

Myers, J. E., & Sweeney, T. J. (2007). Wellness in counseling: An overview (ACAPCD-09). Alexandria, VA: American Counseling Association.

Oishi, S., Diener, E., & Lucas, R.E. (2007). The optimum level of well-being: Can people be too happy? *Perspectives on Psychological Science*, *2*, 346–360.

Padrao, M.J. & Coimbra J.L. (2011). The anorectic dance: Towards a new understanding of inner-experience through psychotherapeutic movement. *American Journal of Dance Therapy*, *33*, 131–147.

Panina, N., & Golovakha, E. (2001). *Tendencies in the development of Ukrainian society (1994–1998), Sociological Indicators*. Kyiv: Institute of Sociology.

Phillips, W.M. (1980). Purpose in life, depression and locus of control. *Journal of Clinical Psychology*, *36*, 661–667.

Rean, A.A. (2001). *Handbook of personality assessment*. Saint-Petersburg: Saint-Petersburg University Press.

Reich, W. (1949). *Character Analysis,* (3rd ed.). New York: Farrar, Straus, and Giroux.

Reker, G. (1977). The purpose-in-life test in an inmate population: An empirical investigation. *Journal of Clinical Psychology*, *33*, 688–693.

Robins, C.J., Ladd, J., Welkowitz, J., Blaney, P.H., Diaz, R., & Kutcher, G. (1994). The Personal Style Inventory: Preliminary validation studies of new measures of sociotropy and autonomy. *Journal of Psychopathology and Behavior Assessment*, *16*, 277–300.

Sammon, S.D., Reznikoff, M., & Geisenger, K.F. (1985). Psychosocial development and stressful life events among religious professionals. *Journal of Personality and Social Psychology*, *48*, 676–687.

Sato, T., & McCann, D. (1998). Individual differences in relatedness and individuality: An exploration of two constructs. *Personality and Individual Differences*, *24*, 847–859.

Schore, A.N. (2001). The effect of a secure attachment relationship on right brain development, affect regulation, and infant mental health. *Infant Mental Health Journal*, *22*, 7-66.

Schore, A.N. (2002). The neurobiology of attachment and early personality organization. *Journal of Prenatal and Perinatal Psychology and Health*, *16*(3), 249-263.

Seber, G.A.F., & Wild, C.J. (2003). *Nonlinear Regression.* Hoboken, New Jersey: John Wiley & Sons, Inc.

Seligman, M.E.P., Rashid, T., & Parks, A. C. (2006). Positive psychotherapy. *American Psychologist*, 61, 774-788.

Sheldon, W. (1940). *Varieties of human physique.* New York, NY: Harper and Row.

Shostrom, E. (1974). *Personal Orientation Inventory: Manual*. San Diego: Edits.

Sin, N.L., & Lyubomirsky, S. (2009). Enhancing well-being and alleviating depressive symptoms with positive psychology interventions: A practice-friendly meta-analysis. *Journal of Clinical Psychology*, *65*, 467–487.

Stark, A. (1987). American Dance Therapy Association, A kinesthetic approach. *Dance Magazine*, *61*, 56–57.

Steel, P., Schnnidt, J., & Shultz, J. (2008). Refining the relationship between personality and subjective well-being. *Psychological Bulletin*, *134***,** 138–161.

Taylor, G.J., Ryan, D.P., & Bagby, R.M. (1985). Toward the development of a new self-report alexithymia scale. *Psychotherapy and Psychosomatics*, *44*, 191–199.

Taylor, M.F., Brice, J., Buck, N., & Prentice-Lane, E. (Eds.) (2009). *British Household Panel Survey User Manual, Volume A: Introduction, technical report and appendices*. Colchester: University of Essex.

Wagner, G.G., Frick, J.R., & Schupp, J. (2007). The German Socio-Economic Panel Study (SOEP). Scope, evolution and enhancements. *Schmollers Jahrbuch*, *127*, 139–169.

Waters, E., Merrick, S., Treboux, D., Crowell, J., & Albersheim, L. (2000). Attachment security in infancy and early adulthood: a twenty-year longitudinal study. *Child Development, 71*, 684–689.
16